\documentclass[showpacs,amssymb,preprint,preprintnumbers,nofootinbib,superscriptaddress]{revtex4}
\usepackage{amsmath}
\usepackage{graphicx}
\usepackage{latexsym}
\usepackage{amsfonts}
\usepackage{url,hyperref}
\usepackage{bm}
\usepackage{textcomp}
\usepackage{color}

\newcommand{\beq}{\begin{equation}}
\newcommand{\eeq}{\end{equation}}
\def\bea{\begin{eqnarray}}
\def\eea{\end{eqnarray}}

\begin{document}

\begin{flushright}
WITS-MITP-004
\end{flushright}

\title{Gravitino fields in Schwarzschild black hole spacetimes}
\author{C.-H. Chen}
\email[Email: ]{899210057@s99.tku.edu.tw}
\affiliation{Department of Physics, Tamkang University, Tamsui, Taipei, Taiwan}
\author{H.~T.~Cho}
\email[Email: ]{htcho@mail.tku.edu.tw}
\affiliation{Department of Physics, Tamkang University, Tamsui, Taipei, Taiwan}
\author{A.~S.~Cornell}
\email[Email: ]{alan.cornell@wits.ac.za}
\affiliation{National Institute for Theoretical Physics, School of Physics and Mandelstam Institute for Theoretical Physics,
University of the Witwatersrand, Johannesburg, Wits 2050, South Africa}
\author{G. Harmsen}
\email[Email: ]{gerhard.harmsen5@gmail.com}
\affiliation{National Institute for Theoretical Physics, School of Physics and Mandelstam Institute for Theoretical Physics,
University of the Witwatersrand, Johannesburg, Wits 2050, South Africa}
\author{Wade~Naylor}
\email[Email: ]{naylor@phys.sci.osaka-u.ac.jp}
\affiliation{International College \& Department of Physics, Osaka University, Toyonaka, Osaka 560-0043, Japan}

\begin{abstract}
The analysis of gravitino fields in curved spacetimes is usually carried out using the Newman-Penrose formalism. In this paper we consider a more direct approach with eigenspinor-vectors on spheres, to separate out the angular parts of the fields in a Schwarzschild background. The radial equations of the corresponding gauge invariant variable obtained are shown to be the same as in the Newman-Penrose formalism. These equations are then applied to the evaluation of the quasinormal mode frequencies, as well as the absorption probabilities of the gravitino field scattering in this background.
\end{abstract}

\date{April 9, 2015}
\maketitle

%
%

\section{Introduction}\label{sec:intro}


\par There has been a lot of interest in understanding the spin-3/2 (or the Rarita-Schwinger \cite{rarsch}) fields after the introduction of the supergravity theory \cite{van}. This spin-3/2 field, or the gravitino, is the superpartner of the graviton. However, the consideration of this field in curved spacetimes, particularly in black hole cases, is relatively sparse. In Refs.~\cite{cassil1} and \cite{cassil2} the gravitino field was analyzed in the Kerr and Kerr-Newman backgrounds. The authors there used the Newman-Penrose formalism \cite{newpen}, following the earlier work of Chandrasekhar on the Dirac field \cite{cha}. The Newman-Penrose formalism is very useful in dealing with perturbations of the black hole in four dimensions, but the extension of the formalism to higher dimensions is not straightforward \cite{gommar}. For this reason we try, in this paper, to implement a novel and more direct approach to considering the gravitino fields in spherically symmetric spacetimes. In this approach we shall use the eigenspinors and the eigenspinor-vectors on spheres to separate out the angular parts of the gravitino fields. We can then obtain the radial equations for the various field components.

\par These field components are gauge dependent in general. The gauge symmetry here is that of supersymmetry. In the Newman-Penrose formalism the radial equations for some scalar potentials are considered, where these potentials are gauge invariant. In order to compare with this formalism we also need to construct gauge invariant combinations out of our field components. The radial equations of these gauge invariant variables can then be used in the analysis of the evolution of spin-3/2 perturbations in spherically symmetry spacetimes, like the Schwarzschild background we concentrate on in this paper. We hope to extend this formalism to higher dimensions in our subsequent works.

\par The study of the evolution of small perturbations in black hole backgrounds is an old and well established subject \cite{koksch,bercar}. Actually, we know that this evolution, at intermediate times is dominated by damped single frequency oscillations. These characteristic oscillations have been termed quasinormal frequencies, and depend only on the parameters characterising the black hole. In this sense we can say that black holes have a characteristic sound \cite{nol}.

\par Note that since the discovery of quasinormal oscillations of black holes, many studies have been done on quasinormal modes (QNMs) of various spin fields and a considerable variety of analytical, semi-analytical and numerical methods have been developed to determine them \cite{chadet,lea,schwil,iyewil,chocor1}. However, an interesting problem is to consider the study of the QNMs of spin-3/2 fields, where the motivation for studying spin-3/2 QNMs from black holes in this paper is two-fold; the first of these being from the theoretical point of view, where the scarcity of work done in four dimensions \cite{pie,cho1} (and in a future work, greater than four dimensions \cite{checho}) is an omission in the literature. Our calculations serve to fill this gap.

\par Secondly, having a complete catalog of all QNMs would be a necessary precursor to eventually studying the emission rates of collider produced, or TeV scale black holes \cite{kan}. Especially given that in super gravity theories the gravitational field is coupled to the massless spin-3/2 Rarita-Schwinger field, that acts as a source of torsion and curvature \cite{dasfre,gripen}. Furthermore, in many supergravity models, the supersymmetric breaking scale is related to the gravitino. As such, the gravitino could be the lightest supersymmetric particle, or even the next to lightest supersymmetric particle (which could live long enough to be directly detectable in collider environments). As such, spin-3/2 fields could have significant effects on phenomenology.

\par Other than the quasinormal modes one is also interested in the greybody factors and the emission cross-sections of Hawking radiation of various spins when black hole evolution in collider environments are studied. The related absorption probabilities of the gravitino scatterings can be readily considered using the radial equation mentioned above.

\par The approach we shall take is to use the WKB method, as it allows a more systematic study of QNMs, as well as the absorption probabilities, than outright numerical methods. We will use this, and previous results with WKB and similar methods \cite{pie}, to compare with the newer method developed by some of us, which can be more efficient in some cases, called the asymptotic iteration method (AIM) \cite{chocor2,doucho,chocor3,chocor1}. In the case of absorption probabilities we shall also study the low energy case using the Unruh method.

\par As such, this paper shall be structured as follows. In the next section we shall use the eigenspinors and the eigenspinor-vectors on the 2-sphere to separate out the angular parts of the gravitino, or the Rarita-Schwinger, equations. In Section III, we construct the gauge-invariant variable of the gravitino field. The corresponding radial equations are compared with those obtained using the Newman-Penrose formalism. In Sections IV and V, the quasinormal mode frequencies and the absorption probabilities of the gravitino field are studied using the radial equation of the gauge invariant variable, respectively. Conclusions and discussions are presented in the last section. Some properties of the eigen-spinors and the eigen-spinor vectors relevant to our discussions are given in the Appendix.

%
%

\section{Massless Rarita-Schwinger equation}\label{sec:RSEquation}

\par We start with the massless Rarita-Schwinger equation
\begin{equation}
\gamma^{\mu\nu\alpha}\nabla_{\nu}\psi_{\alpha}=0\; , \label{RSequation}
\end{equation}
where we have used the notation
\begin{eqnarray}
\gamma^{\mu\nu\alpha}=\gamma^{[\mu}\gamma^{\nu}\gamma^{\alpha]}
=\gamma^{\mu}\gamma^{\nu}\gamma^{\alpha}-\gamma^{\mu}g^{\nu\alpha}+\gamma^{\nu}g^{\mu\alpha}-\gamma^{\alpha}g^{\mu\nu}\; .
\end{eqnarray}
Here we consider the Rarita-Schwinger field in a Schwarzschild black hole background with the metric
\begin{eqnarray}
ds^{2}=-f(r)dt^{2}+f(r)^{-1}dr^{2}+r^{2}d\bar{\Omega}^{2}_{2} \; ,
\end{eqnarray}
where $f(r)=1-2M/r$. Note that from now on an overbar represents quantities on the 2-sphere.

\par Our choice of Dirac matrices are as follows:
\begin{eqnarray}
\gamma^{0}&=&i\sigma^{3}\otimes\mathbf{1}\Rightarrow\gamma^{t}=\frac{1}{\sqrt{f}}(i\sigma^{3}\otimes\mathbf{1})\; ,\nonumber\\
\gamma^{i}&=&\sigma^{1}\otimes\bar{\gamma}^{i}\Rightarrow\gamma^{\theta_{i}}=\frac{1}{r}(\sigma^{1}\otimes\bar{\gamma}^{\theta_{i}})\; ,\nonumber\\
\gamma^{3}&=&\sigma^{2}\otimes\mathbf{1}\Rightarrow\gamma^{r}=\sqrt{f}(\sigma^{2}\otimes\mathbf{1})\; ,
\end{eqnarray}
where $\mathbf{1}$ is the $2\times 2$ unit matrix, $\sigma^{i}$ $(i=1,2,3)$ are the Pauli matrices, and $\bar{\gamma}^{i}$ $(i=1,2)$ are the Dirac matrices for a 2-sphere. The corresponding non-zero spinor connection components are
\begin{eqnarray}
&&\Gamma_{t}=-\frac{f'}{4}(\sigma^{1}\otimes\mathbf{1})\; ,\ \ \ \Gamma_{r}=0\; ,\nonumber\\
&&\Gamma_{\theta_{i}}=\mathbf{1}\otimes\bar{\Gamma}_{\theta_{i}}+\frac{\sqrt{f}}{2}(i\sigma^{3}\otimes\bar{\gamma}_{\theta_{i}})\; .
\end{eqnarray}

\par Now we can work on Eq.~(\ref{RSequation}). First, the product of Dirac gamma functions are found to be
\begin{eqnarray}
\gamma^{t\theta_{i}r}&=&-\frac{1}{r}\left(\mathbf{1}\otimes\bar{\gamma}^{\theta_{i}}\right)\; ,\ \ \
\gamma^{t\theta_{i}\theta_{j}}=\frac{1}{r^{2}\sqrt{f}}\left(i\sigma^{3}\otimes\bar{\gamma}^{\theta_{i}\theta_{j}}\right)\; ,\nonumber\\
\gamma^{r\theta_{i}\theta_{j}}&=&\frac{\sqrt{f}}{r^{2}}\left(\sigma^{2}\otimes\bar{\gamma}^{\theta_{i}\theta_{j}}\right)\; ,
\end{eqnarray}
where $\bar{\gamma}^{\theta_{i}\theta_{j}}=\bar{\gamma}^{[\theta_{i}}\bar{\gamma}^{\theta_{j}]}$.
Then we separate the fields into their $r$-$t$ and angular parts. Since $\psi_{t}$ and $\psi_{r}$ behave like spinors on the 2-sphere, we can write
\begin{equation}
\psi_{t}=\phi_{t}\otimes\bar{\psi}_{\lambda}\; ,\ \ \ \psi_{r}=\phi_{r}\otimes\bar{\psi}_{\lambda}\; ,\label{psit}
\end{equation}
where $\bar{\psi}_{\lambda}$ is a eigenspinor with eigenvalue $i\bar{\lambda}$. For completeness the eigenspinors on the 2-sphere are briefly described in the Appendix. $\phi_{t}$ and $\phi_{r}$ are functions of $r$ and $t$. Since the Schwarzschild metric is static, one can consider one frequency at a time and assume the time dependence to be $e^{-i\omega t}$.

\par $\psi_{\theta_{i}}$ behaves like a spinor-vector on the 2-sphere. We can therefore write
\begin{equation} \psi_{\theta_{i}}=\phi_{\theta}^{(1)}\otimes\bar{\nabla}_{\theta_{i}}\bar{\psi}_{\lambda}
+\phi_{\theta}^{(2)}\otimes\bar{\gamma}_{\theta_{i}}\bar{\psi}_{\lambda}\; ,\label{psitheta}
\end{equation}
where we have represented the two sets of eigenspinor-vectors on the 2-sphere by $\bar{\nabla}_{\theta_{i}}\bar{\psi}_{\lambda}$ and $\bar{\gamma}_{\theta_{i}}\bar{\psi}_{\lambda}$. Some properties of these eigenspinor-vectors are also presented in the Appendix.


\par The equations of motion for $\phi_{t}$, $\phi_{r}$ and $\phi_{\theta_{i}}$ can be derived by analyzing Eq.~(\ref{RSequation}). For simplicity, we shall take the Weyl or temporal gauge such that $\phi_{t}=0$. Gauge symmetry of the massless Rarita-Schwinger equation will be discussed in more detail in the next section.

\par For $\mu=t$ in Eq.~(\ref{RSequation}), we have
\begin{eqnarray}
&&\gamma^{t\nu\alpha}\nabla_{\nu}\psi_{\alpha}=0\nonumber\\
&\Rightarrow&\gamma^{t\theta_{i}r}\left(\partial_{\theta_{i}}\psi_{r}
+\Gamma_{\theta_{i}}\psi_{r}-\partial_{r}\psi_{\theta_{i}}\right)
+\gamma^{t\theta_{i}\theta_{j}}\left(\partial_{\theta_{i}}\psi_{\theta_{j}}
+\Gamma_{\theta_{i}}\psi_{\theta_{j}}\right)=0\; .\label{tRSeqn}
\end{eqnarray}
Using the definitions in Eqs.~(\ref{psit}) and (\ref{psitheta}), Eq.~(\ref{tRSeqn}) can be simplified to
\begin{eqnarray}
\Bigg[-i\,\bar{\lambda}\phi_{r}-\sqrt{f}(i\sigma^{3})\phi_{r}+i\,\bar{\lambda}\partial_{r}\phi_{\theta}^{(1)}
+\frac{i\,\bar{\lambda}}{2r}\phi_{\theta}^{(1)}-\frac{1}{2r\sqrt{f}}(i\sigma^{3})\phi_{\theta}^{(1)}&&\nonumber\\
+2\partial_{r}\phi_{\theta}^{(2)}+\frac{i\,\bar{\lambda}}{r\sqrt{f}}(i\sigma^{3})\phi_{\theta}^{(2)}
+\frac{1}{r}\phi_{\theta}^{(2)}\Bigg]\otimes\bar{\psi}&=&0\; .
\end{eqnarray}
Hence, we have the first equation involving $\phi_{r}$, $\phi_{\theta}^{(1)}$ and $\phi_{\theta}^{(2)}$,
\begin{eqnarray}
\left(i\,\bar{\lambda}\,\partial_{r}
+\frac{i\,\bar{\lambda}}{2r}-\frac{1}{2r\sqrt{f}}(i\sigma^{3})\right)\phi_{\theta}^{(1)}
+\left(2\,\partial_{r}+\frac{i\,\bar{\lambda}}{r\sqrt{f}}(i\sigma^{3})
+\frac{1}{r}\right)\phi_{\theta}^{(2)}&&\nonumber\\
-\left(i\,\bar{\lambda}+\sqrt{f}(i\sigma^{3})\right)\phi_{r}&=&0\; .\label{eom1}
\end{eqnarray}

\par Similarly for $\mu=r$ we have from Eq.~(\ref{RSequation}),
\begin{eqnarray}
&&\gamma^{r\nu\alpha}\nabla_{\nu}\psi_{\alpha}=0\nonumber\\
&\Rightarrow&\gamma^{r\theta_{i}t}\left(\partial_{\theta_{i}}\psi_{t}
+\Gamma_{\theta_{i}}\psi_{t}-\partial_{t}\psi_{\theta_{i}}-\Gamma_{t}\psi_{\theta_{i}}\right)
+\gamma^{r\theta_{i}\theta_{j}}\left(\partial_{\theta_{i}}\psi_{\theta_{j}}+\Gamma_{\theta_{i}}\psi_{\theta_{j}}\right)=0\; .
\label{rRSeqn}
\end{eqnarray}
Again using Eqs.~(\ref{psit}) and (\ref{psitheta}), Eq.~(\ref{rRSeqn}) becomes
\begin{eqnarray}
&&\Bigg[-\frac{i\,\bar{\lambda}}{\sqrt{f}}\partial_{t}\phi_{\theta}^{(1)}+\frac{i\,\bar{\lambda}\sqrt{f}}{2r}\sigma^{1}\phi_{\theta}^{(1)}
+\frac{i\,\bar{\lambda}f'}{4\sqrt{f}}\sigma^{1}\phi_{\theta}^{(1)}-\frac{1}{2r}\sigma^{2}\phi_{\theta}^{(1)}\nonumber\\
&&\ \ -\frac{2}{\sqrt{f}}\partial_{t}\phi_{\theta}^{(2)}+\frac{\sqrt{f}}{r}\sigma^{1}\phi_{\theta}^{(2)}
+\frac{f'}{2\sqrt{f}}\sigma^{1}\phi_{\theta}^{(2)}+\frac{i\,\bar{\lambda}}{r}\sigma^{2}\phi_{\theta}^{(2)}\Bigg]\otimes\bar{\psi}=0\; .
\end{eqnarray}
Note that we have taken the Weyl gauge: $\phi_{t}=0$. Hence, we have the second equation of motion
\begin{eqnarray}
&&\left(-\frac{i\,\bar{\lambda}}{\sqrt{f}}\partial_{t}+\frac{i\,\bar{\lambda}\sqrt{f}}{2r}\sigma^{1}
+\frac{i\,\bar{\lambda}f'}{4\sqrt{f}}\sigma^{1}-\frac{1}{2r}\sigma^{2}\right)\phi_{\theta}^{(1)}\nonumber\\
&&\ \ \ \ \ +\left(-\frac{2}{\sqrt{f}}\partial_{t}+\frac{\sqrt{f}}{r}\sigma^{1}
+\frac{f'}{2\sqrt{f}}\sigma^{1}+\frac{i\,\bar{\lambda}}{r}\sigma^{2}\right)\phi_{\theta}^{(2)}=0\; .\label{eom2}
\end{eqnarray}

\par Lastly, for $\mu=\theta_{i}$ in Eq.~(\ref{RSequation}), we have
\begin{eqnarray}
&&\gamma^{\theta_{i}\nu\alpha}\nabla_{\nu}\psi_{\alpha}=0\nonumber\\
&\Rightarrow&\gamma^{\theta_{i}tr}\left(\partial_{t}\psi_{r}+\Gamma_{t}\psi_{r}\right)+
\gamma^{\theta_{i}\theta_{j}r}\left(\partial_{\theta_{j}}\psi_{r}+\Gamma_{\theta_{j}}\psi_{r}-\partial_{r}\psi_{\theta_{j}}\right)
+\gamma^{\theta_{i}\theta_{j}t}\left(-\partial_{t}\psi_{\theta_{j}}-\Gamma_{t}\psi_{\theta_{j}}\right)=0\nonumber\\
&\Rightarrow&\Bigg[-\frac{\sqrt{f}}{r}\sigma^{2}\phi_{r}+\left(\frac{1}{r\sqrt{f}}(i\sigma^{3})\partial_{t}
+\frac{\sqrt{f}}{r}\sigma^{2}\partial_{r}+\frac{f'}{4r\sqrt{f}}\sigma^{2}\right)\phi_{\theta}^{(1)}\Bigg]
\otimes\bar{\nabla}^{\theta_{i}}\bar{\psi}\nonumber\\
&&+\Bigg[\left(\partial_{t}-\frac{f}{2r}\sigma^{1}-\frac{f'}{4}\sigma^{1}+\frac{i\bar{\lambda}\sqrt{f}}{r}\sigma^{2}\right)\phi_{r}
-\left(\frac{i\bar{\lambda}}{r\sqrt{f}}(i\sigma^{3})\partial_{t}+\frac{i\bar{\lambda}\sqrt{f}}{r}\sigma^{2}\partial_{r}
+\frac{i\bar{\lambda}f'}{4r\sqrt{f}}\sigma^{2}\right)\phi_{\theta}^{(1)}\nonumber\\
&&\ \ \ \ \ \ \ -\left(\frac{1}{r\sqrt{f}}(i\sigma^{3})\partial_{t}+\frac{\sqrt{f}}{r}\sigma^{2}\partial_{r}+\frac{f'}{4r\sqrt{f}}\sigma^{2}\right)\phi_{\theta}^{(2)}\Bigg]
\otimes\bar{\gamma}^{\theta_{i}}\bar{\psi}=0\; .
\end{eqnarray}
Therefore we have two more equations of motion
\begin{eqnarray}
&&-\frac{\sqrt{f}}{r}\sigma^{2}\phi_{r}+\left(\frac{1}{r\sqrt{f}}(i\sigma^{3})\partial_{t}
+\frac{\sqrt{f}}{r}\sigma^{2}\partial_{r}+\frac{f'}{4r\sqrt{f}}\sigma^{2}\right)\phi_{\theta}^{(1)}=0\; ,\label{eom3}
\end{eqnarray}
and
\begin{eqnarray}
\left(\partial_{t}-\frac{f}{2r}\sigma^{1}-\frac{f'}{4}\sigma^{1}+\frac{i\bar{\lambda}\sqrt{f}}{r}\sigma^{2}\right)\phi_{r}
-\left(\frac{i\bar{\lambda}}{r\sqrt{f}}(i\sigma^{3})\partial_{t}+\frac{i\bar{\lambda}\sqrt{f}}{r}\sigma^{2}\partial_{r}
+\frac{i\bar{\lambda}f'}{4r\sqrt{f}}\sigma^{2}\right)\phi_{\theta}^{(1)}&&\nonumber\\
-\left(\frac{1}{r\sqrt{f}}(i\sigma^{3})\partial_{t}+\frac{\sqrt{f}}{r}\sigma^{2}\partial_{r}+\frac{f'}{4r\sqrt{f}}\sigma^{2}\right)\phi_{\theta}^{(2)}&=&0\; . \nonumber \\
&&\label{eom4}
\end{eqnarray}

\par In summary we have four equations of motion, Eqs.~(\ref{eom1}), (\ref{eom2}), (\ref{eom3}), and (\ref{eom4}) for three functions $\phi_{r}$, $\phi_{\theta}^{(1)}$, and $\phi_{\theta}^{(2)}$. One can show that of these four equations, only three of them are independent. On the other hand, as we shall see below, the functions $\phi_{r}$, $\phi_{\theta}^{(1)}$, and $\phi_{\theta}^{(2)}$ are not gauge-invariant. Therefore, in the next section we shall consider the gauge symmetry of the Rarita-Schwinger equation in detail. Then we shall construct the appropriate gauge-invariant variables and derive their corresponding radial equations of motion.

%
%

\section{Gauge-invariant variable}

\par In this section we consider the gauge freedom for Eq.~(\ref{RSequation}). Consider the transformation of $\psi_{\mu}$,
\begin{equation}
\psi'_{\mu}=\psi_{\mu}+\nabla_{\mu}\varphi \; ,\label{transform}
\end{equation}
where $\varphi$ is a Dirac spinor. Plugging this into Eq.~(\ref{RSequation}), we must have
\begin{equation}
\gamma^{\mu\nu\alpha}\nabla_{\nu}\nabla_{\alpha}\varphi=0
\end{equation}
for the equation to possess this gauge symmetry. Now,
\begin{eqnarray}
\gamma^{\mu\nu\alpha}\nabla_{\nu}\nabla_{\alpha}\varphi
&=&\frac{1}{2}\gamma^{\mu\nu\alpha}[\nabla_{\nu},\nabla_{\alpha}]\varphi\nonumber\\
&=&\frac{1}{8}\gamma^{\mu\nu\alpha}R_{\nu\alpha\rho\sigma}\gamma^{\rho}\gamma^{\sigma}\varphi \; .\label{gauge}
\end{eqnarray}
Note that to obtain the result for the commutator, $[\nabla_{\nu},\nabla_{\alpha}]$, we have assumed that there are only gravitational interactions. For example, for a charged black hole background there should be extra terms proportional to the electromagnetic field strength.

\par To simplify this expression in Eq.~(\ref{gauge}) further we need the following identity. From the symmetry of the Riemann tensor,
\begin{eqnarray}
\gamma^{\mu}\gamma^{\nu}\gamma^{\alpha}(R_{\mu\nu\alpha\beta}+R_{\nu\alpha\mu\beta}+R_{\alpha\mu\nu\beta})=0
&\Rightarrow&3(\gamma^{\mu}\gamma^{\nu}\gamma^{\alpha}R_{\mu\nu\alpha\beta}+2\gamma^{\alpha}R_{\alpha\beta})=0\nonumber\\
&\Rightarrow&\gamma^{\mu}\gamma^{\nu}\gamma^{\alpha}R_{\mu\nu\alpha\beta}=-2\gamma^{\alpha}R_{\alpha\beta}\; .
\end{eqnarray}
Also we have
\begin{equation}
\gamma^{\mu}\gamma^{\nu}\gamma^{\alpha}\gamma^{\beta}R_{\mu\nu\alpha\beta}=-2R\; .
\end{equation}
Using these identities Eq.~(\ref{gauge}) becomes
\begin{eqnarray}
\frac{1}{8}\gamma^{\mu\nu\alpha}R_{\nu\alpha\rho\sigma}\gamma^{\rho}\gamma^{\sigma}\varphi
&=&\frac{1}{8}\left(\gamma^{\mu}\gamma^{\nu}\gamma^{\alpha}-\gamma^{\mu}g^{\nu\alpha}+\gamma^{\nu}g^{\mu\alpha}-\gamma^{\alpha}g^{\mu\nu}\right)
\gamma^{\rho}\gamma^{\sigma}R_{\nu\alpha\rho\sigma}\varphi\nonumber\\
&=&\frac{1}{4}(2\gamma^{\alpha}{R_{\alpha}}^{\mu}-\gamma^{\mu}R)\varphi \; .
\end{eqnarray}
This is zero for Ricci flat spacetimes like the four dimensional Schwarzschild spacetime that we look at below. However, even for de Sitter and anti-de Sitter spacetimes it does not vanish, so we need to modify the covariant derivative in those cases in order to respect the gauge symmetry.

\par Consider the gauge transformations on $\phi_{t}$ and $\phi_{r}$. Take $\varphi=\phi\otimes\bar{\psi}_{\lambda}$, then Eq.~(\ref{transform}) becomes
\begin{eqnarray}
\psi'_{t}=\psi_{t}+\nabla_{t}\varphi
&\Rightarrow&\phi'_{t}=\phi_{t}+\partial_{t}\phi-\frac{f'}{4}\sigma^{1}\phi\; , \\
\psi'_{r}=\psi_{r}+\nabla_{r}\varphi
&\Rightarrow&\phi'_{r}=\phi_{r}+\partial_{r}\phi\; .
\end{eqnarray}
The gauge transformation of the angular components of $\psi_{\mu}$ is more complicated.
\begin{eqnarray}
&&\psi'_{\theta_{i}}=\psi_{\theta_{i}}+\nabla_{\theta_{i}}\varphi\nonumber\\
&\Rightarrow&\phi'^{(1)}_{\theta}\otimes\bar{\nabla}_{\theta_{i}}\bar{\psi}_{\lambda}
+\phi'^{(2)}\otimes\bar{\gamma}_{\theta_{i}}\bar{\psi}_{\lambda}=\left(\phi_{\theta}^{(1)}+\phi\right)\otimes\bar{\nabla}_{\theta_{i}}
\bar{\psi}_{\lambda}+\left(\phi_{\theta}^{(2)}+\frac{\sqrt{f}}{2}(i\sigma^{3})\phi\right)\otimes\bar{\gamma}_{\theta_{i}}\bar{\psi}_{\lambda}\nonumber\\
&\Rightarrow&\phi'^{(1)}_{\theta}=\phi_{\theta}^{(1)}+\phi\ \ \ ;\ \ \ \phi'^{(2)}=\phi_{\theta}^{(2)}+\frac{\sqrt{f}}{2}(i\sigma^{3})\phi\; .
\end{eqnarray}
We see that indeed $\phi_{t}$, $\phi_{r}$, $\phi_{\theta}^{(1)}$, and $\phi_{\theta}^{(2)}$ are not gauge invariant.

\par Considering the gauge transformations of these functions, one can construct gauge-invariant combinations. For example, a simple one would be
\begin{eqnarray}
\Phi=-\frac{\sqrt{f}}{2}(i\sigma^{3})\phi_{\theta}^{(1)}+\phi_{\theta}^{(2)}\; .
\end{eqnarray}
In the following we shall look for the radial equation for $\Phi$ and then we shall show that it is the same as the one obtained in Refs. \cite{cassil1} and \cite{cassil2}.

\par Now we go back to the equations of motion in Eqs.~(\ref{eom1}), (\ref{eom2}), (\ref{eom3}), and (\ref{eom4}). Only three of them are independent and we shall consider Eqs.~(\ref{eom1}), (\ref{eom2}), and (\ref{eom3}). From these three equations we can derive the equation of motion for $\Phi$,
\begin{eqnarray}
\left(\sqrt{f}+\bar{\lambda}\sigma^{3}\right)\Bigg[-\sigma^{1}\partial_{t}\Phi+\frac{f'}{4}\Phi
&+&\frac{\sqrt{f}}{2r}(\sqrt{f}-\bar{\lambda}\sigma^{3})\Phi\Bigg] \nonumber \\
&=&\left(\sqrt{f}-\bar{\lambda}\sigma^{3}\right)\left[f\partial_{r}\Phi+\frac{\sqrt{f}}{2r}(\sqrt{f}-\bar{\lambda}\sigma^{3})\Phi\right]\; .\label{Phieqn}
\end{eqnarray}
Note that in arriving at this form we have used the fact that $f=1-2M/r$.

\par Next we try to transform the radial equation, Eq.~(\ref{Phieqn}), into the Schr\"odinger form. To do that we decompose the equation into its components,
\begin{eqnarray}
\Phi=\left(
\begin{array}{c}
\Phi_{1} \\ \Phi_{2}
\end{array}
\right)\; ,
\end{eqnarray}
and Eq.~(\ref{Phieqn}) becomes two equations
\begin{eqnarray}
&&\left(\frac{\sqrt{f}-\bar{\lambda}}{\sqrt{f}+\bar{\lambda}}\right)f\partial_{r}\Phi_{1}-\frac{f'}{4}\Phi_{1}
-\frac{\sqrt{f}}{r}\left(\frac{\sqrt{f}-\bar{\lambda}}{\sqrt{f}+\bar{\lambda}}\right)\bar{\lambda}\Phi_{1}
=i\omega\Phi_{2}\; ,\label{Phi1}\\
&&\left(\frac{\sqrt{f}+\bar{\lambda}}{\sqrt{f}-\bar{\lambda}}\right)f\partial_{r}\Phi_{2}-\frac{f'}{4}\Phi_{2}
+\frac{\sqrt{f}}{r}\left(\frac{\sqrt{f}+\bar{\lambda}}{\sqrt{f}-\bar{\lambda}}\right)\bar{\lambda}\Phi_{2}
=i\omega\Phi_{1}\; ,\label{Phi2}
\end{eqnarray}
where we assume the time dependence of the fields is $e^{-i\omega t}$.

\par To further simplify the equations we make the following transformation,
\begin{eqnarray}
\tilde{\Phi}_{1}=\left(\frac{f^{1/4}}{\sqrt{f}+\bar{\lambda}}\right)\Phi_{1}\ \ \ ;
\ \ \ \tilde{\Phi}_{2}=\left(\frac{f^{1/4}}{\sqrt{f}-\bar{\lambda}}\right)\Phi_{2}\; .
\end{eqnarray}
Then Eqs.~(\ref{Phi1}) and (\ref{Phi2}) become
\begin{eqnarray}
\left(\frac{d}{dr_{*}}-W\right)\tilde{\Phi}_{1}=i\omega\tilde{\Phi}_{2}\ \ \ ;\ \ \ \left(\frac{d}{dr_{*}}+W\right)\tilde{\Phi}_{2}=i\omega\tilde{\Phi}_{1}\; , \label{susyform}
\end{eqnarray}
with $d/dr_{*}=f\,d/dr$ and
\begin{eqnarray}
W=\frac{|\bar{\lambda}|\sqrt{f}}{r}\left(\frac{\bar{\lambda}^{2}-1}{\bar{\lambda}^{2}-f}\right)
=\frac{\left(j-\frac{1}{2}\right)\left(j+\frac{1}{2}\right)\left(j+\frac{3}{2}\right)\sqrt{f}}
{r\left[\left(j+\frac{1}{2}\right)^{2}-f\right]}\; ,
\end{eqnarray}
where for convenience we have written $|\bar{\lambda}|=j+1/2$ with $j=3/2,5/2,\dots$.

\par
\begin{figure}[!]
\centering
\includegraphics[height=80mm]{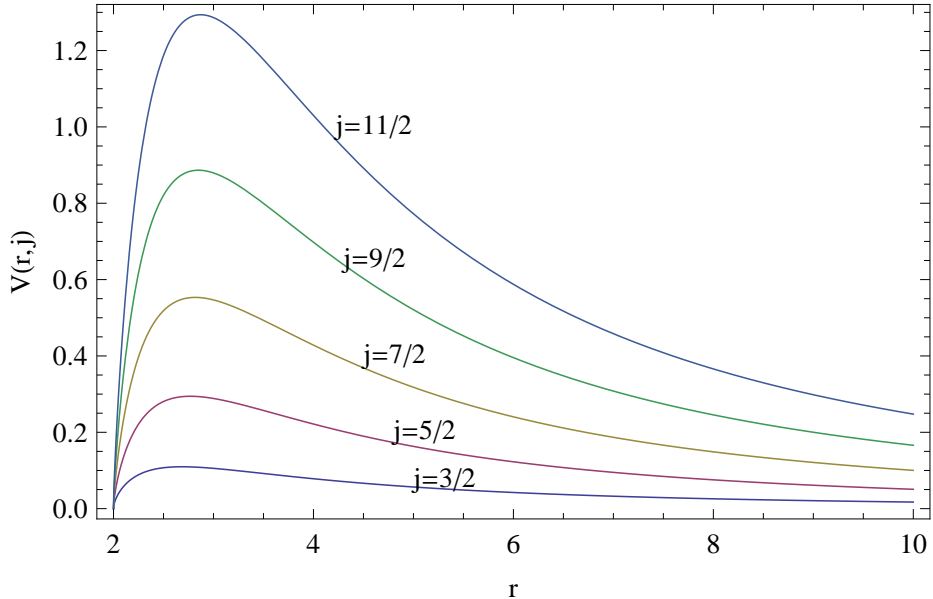}
\caption{\it Effective potentials of the gravitino field with $j=3/2$ to $j=11/2$.}
\label{fig1}
\end{figure}

 Eq.~(\ref{susyform}) is in the form of the so-called SUSY quantum mechanics \cite{cookha} with $W$ being the superpotential. Now we can construct two Schr\"odinger-like equations,
\begin{eqnarray}
-\frac{d^{2}}{dr_{*}^{2}}\tilde{\Phi}_{1}+V_{1}\tilde{\Phi}_{1}
=\omega^{2}\tilde{\Phi}_{1}\ \ \ ;\ \ \ -\frac{d^{2}}{dr_{*}^{2}}\tilde{\Phi}_{2}+V_{2}\tilde{\Phi}_{2}
=\omega^{2}\tilde{\Phi}_{2}\; , \label{SL}
\end{eqnarray}
where the isospectral potentials are
\begin{eqnarray}
V_{1,2}=\pm f\frac{dW}{dr}+W^{2}\; .
\end{eqnarray}
More explicitly,
\begin{eqnarray}
V_{1,2}&=&\frac{\left(j-\frac{1}{2}\right)\left(j+\frac{1}{2}\right)\left(j+\frac{3}{2}\right)\sqrt{1-\frac{2M}{r}}}
{r^{2}\left[\left(j-\frac{1}{2}\right)\left(j+\frac{3}{2}\right)+\frac{2M}{r}\right]^{2}}\times\nonumber\\
&&\ \ \ \left[\pm\frac{2M^{2}}{r^{2}}
+\left(j-\frac{1}{2}\right)\left(j+\frac{3}{2}\right)\left(\left(j+\frac{1}{2}\right)\sqrt{1-\frac{2M}{r}}\pm\frac{3M}{r}\mp 1\right)\right]\; .\label{V}
\end{eqnarray}
$V_{1}$ is the same potential obtained in \cite{cassil1,cassil2}. To see the behavior of $V_{1}$ for various values of $j$ we have plotted them in Fig.~\ref{fig1} for $j=3/2$ to $11/2$. They are similar to the barrier potentials of perturbations of other spins \cite{nol,cho2}.

%
%
%

\section{Quasinormal modes}\label{sec:QNM}

In this section we consider the QNMs corresponding to the potential $V_{1}$ for the gravitino in a Schwarzschild spacetime. For notational simplicity, we shall write $V$ instead of $V_{1}$. The calculation of the QNM frequencies was pioneered by Chandrasekhar and Detweiler \cite{chadet} using numerical methods. Due to the particular boundary conditions for a QNM, the numerical code is not easy to implement. In order to obtain the QNM frequencies in a more efficient manner, other semi-analytical methods have been devised, notably the continued fraction \cite{lea} and WKB methods \cite{schwil,iyewil}.

\par Up until recently, the WKB method has perhaps been the most popular method to evaluate the black hole QNM frequencies. The method was put forth by Schutz and Will \cite{schwil}, then extended to third order by Iyer and Will \cite{iyewil}, and even to sixth order by Konoplya \cite{kon}. In fact, the lowest order in the WKB approximation can be viewed as the large angular momentum limit \cite{chocor4}, that is, the large $j$ limit in the potential of Eq.~(\ref{V}). In this limit, the QNM frequencies are given by the formula
\begin{eqnarray}
\omega^{2}\approx V_{0}-i(n+\frac{1}{2})(-2V''_{0})^{1/2}\; ,
\end{eqnarray}
where $n=0,1,2,\dots$ is the mode number and $V_{0}$ is the maximum of the potential. For $j\rightarrow\infty$, the potential becomes
\begin{eqnarray}
V|_{j\rightarrow\infty}\approx \frac{j^{2}}{r^{2}}\left(1-\frac{2}{r}\right).
\end{eqnarray}
Note that from here on we shall take $M=1$. The maximum of the potential in this limit is located at $r=3$. Hence,
\begin{eqnarray}
V_{0}|_{j\rightarrow\infty}=\frac{j^{2}}{27}\ \ \ ;\ \ \ V''_{0}|_{j\rightarrow\infty}=-\frac{2j^{2}}{729}\; .
\end{eqnarray}
The QNM frequencies in this limit are therefore
\begin{eqnarray}
\omega|_{j\rightarrow\infty}=\frac{1}{3\sqrt{3}}\left[j-i(n+\frac{1}{2})\right]\; .
\end{eqnarray}

\begin{table}
\caption{\it Low-lying ($n\leq l$, with $l=j-3/2$) gravitino quasinormal mode frequencies using the WKB and AIM methods.\label{table1}}
\begin{tabular}{| l | l || l | l || l || l |}
\hline
& & \multicolumn{2}{c ||}{WKB}  & \multicolumn{1}{c||}{AIM} \\
\hline
$l$ & $n$ & 3rd Order &6th Order  & 150 iterations \\
\hline
0 & 0 &  0.3087 - 0.0902i  & 0.3113 - 0.0902i   & 0.3108 - 0.0899i  \\
\hline
1 & 0 &  0.5295 - 0.0938i  & 0.5300 - 0.0938i   & 0.5301 - 0.0937i \\
1 & 1 &  0.5103 - 0.2858i  & 0.5114 - 0.2854i   & 0.5119 - 0.2863i \\
\hline
2 & 0 &  0.7346 - 0.0949i  &  0.7348 - 0.0949i  & 0.7348 - 0.0949i \\
2 & 1 &  0.7206 - 0.2870i  &  0.7210 - 0.2869i  & 0.7211 - 0.2871i \\
2 & 2 &  0.6960 - 0.4844i  &  0.6953 - 0.4855i  & 0.6892 - 0.4834i\\
\hline
3 & 0 &  0.9343 - 0.0954i  &  0.9344 - 0.0954i  & 0.9344 - 0.0954i \\
3 & 1 &  0.9233 - 0.2876i  &  0.9235 - 0.2876i  & 0.9235 - 0.2876i \\
3 & 2 &  0.9031 - 0.4835i  &  0.9026 - 0.4840i  & 0.9026 - 0.4840i \\
3 & 3 &  0.8759 - 0.6835i  & 0.8733 - 0.6870i   & 0.8733 - 0.6870i \\
\hline
4 & 0 &  1.1315 - 0.0956i  &  1.1315 - 0.0956i  & 1.1315 - 0.0956i \\
4 & 1 &  1.1224 - 0.2879i  &  1.1225 - 0.2879i  & 1.1225 - 0.2879i \\
4 & 2 &  1.1053 - 0.4828i  &  1.1050 - 0.4831i  & 1.1050 - 0.4831i \\
4 & 3 &  1.0817 - 0.6812i  &  1.0798 - 0.6830i  & 1.0798 - 0.6830i  \\
4 & 4 &  1.0530 - 0.8828i  &  1.0485 - 0.8891i  & 1.0485 - 0.8891i\\
\hline
5 & 0 &  1.3273 - 0.0958i  &  1.3273 - 0.0958i  & 1.3273 - 0.0958i \\
5 & 1 &  1.3196 - 0.2881i  &  1.3196 - 0.2881i  & 1.3196 - 0.2881i \\
5 & 2 &  1.3048 - 0.4824i  &  1.3045 - 0.4826i  & 1.3046 - 0.4826i \\
5 & 3 &  1.2839 - 0.6795i  &  1.2826 - 0.6805i  & 1.2826 - 0.6805i \\
5 & 4 &  1.2582 - 0.8794i  &  1.2548 - 0.8832i  & 1.2548 - 0.8832i \\
5 & 5 &  1.2284 - 1.0821i  &  1.2221 - 1.0915i  & 1.2221 - 1.0915i \\
\hline
\end{tabular}
\end{table}

\par To obtain the QNM frequencies in small values of $j$, especially the fundamental ones, it is necessary to go higher orders in the WKB approximation. Using the formula put forth in Ref. \cite{iyewil} for third order and in Ref. \cite{kon} for sixth order of the WKB approximation, we have evaluated the frequencies for $j=3/2$ to $13/2$, and with $n\leq l$ ($l=j-3/2$) since the WKB approximation is known to be accurate for low-lying modes. These values are listed in Table~\ref{table1} and can be compared with the ones given recently in Ref. \cite{pie}.

\par In a series of papers \cite{chocor2,doucho,chocor3,chocor1} we have developed a new semi-analytic method to evaluate the QNM frequencies called the asymptotic iteration method (AIM) as an alternative to the WKB approach. To implement the AIM approach in Eq.~(\ref{SL}), we first make a coordinate transformation $y^{2}=1-1/r$, so that the region $1<r<\infty$ is now in a compact one $0<y<1$. Then we extract the asymptotic behaviors of $\tilde{\Phi}_{1}$ and write
\begin{eqnarray}
\tilde{\Phi}_{1}=y^{-i\omega}(1-y^{2})^{-2i\omega}e^{2i\omega/(1-y^{2})}\chi(y)\; .
\end{eqnarray}
The function $\chi(y)$ satisfies the equation
\begin{eqnarray}
\chi''(y)=\lambda_{0}\chi'(y)+s_{0}(y)\; ,\label{chieqn}
\end{eqnarray}
where
\begin{eqnarray}
\lambda_{0}&=&\frac{8i\omega(1-4y^{2}+2y^{4})}{y(1-y^{2})^{2}}-\frac{1-5y^{2}}{y(1-y^{2})}\; ,\\
s_{0}&=&-\frac{64\omega^{2}(2-y^{2})}{(1-y^{2})^{2}}-\frac{32i\omega}{1-y^{2}}\nonumber\\
&&\ \ +\frac{(2j-1)(2j+1)(2j+3)[(2j+1)-2y]^{2}[1+(2j+1)y+y^{2}]}
{y(1-y^{2})^{2}[(1-4j-4j^{2})+4y^{2}-2y^{4}]^{2}}\; .
\end{eqnarray}
The AIM procedure can now be applied to Eq.~(\ref{chieqn}) and the result obtained after 150 iterations are tabulated in Table~\ref{table1}. One can see that the results are definitely better than the third order WKB and are comparable with the sixth order WKB and the prony results in Ref. \cite{pie}.

%
%

\section{Absorption probabilities}

\par For the low energy regime analytic formula of the absorption probability in the scattering of gravitinos can be obtained using the Unruh approximation method \cite{unr}. On the other hand, for the more general energy regimes we shall use the WKB approximation, especially the one developed by Iyer and Will \cite{iyewil}.

\par To implement the Unruh method we begin with the potential in Eq.~(\ref{V}). We shall consider three regions: (I) Near horizon where $f(r)\rightarrow 0$, (II) Central region where the potential is much larger than the energy with $V(r)\gg\omega$, and (III) Far from the black hole where $f(r)\rightarrow 1$. Approximated solutions are obtained for these regions separately and then the unknown coefficients are determined by matching these solutions at the boundary regions.

\par In the first region, $f(r)\rightarrow0$, Eq.~(\ref{SL}) becomes
\begin{equation}
\left(\frac{d^{2}}{dr_{*}^{2}}+\omega ^{2}\right)\tilde{\Phi}_{\rm I}=0\; .
\end{equation}
With the ingoing boundary condition near the event horizon, the solution goes like
\begin{equation}
\tilde{\Phi}_{\rm{I}}=A_{\rm{I}}e^{-i\omega r_{*}}.
\end{equation}

\par Next, in the region $V(r)\gg\omega$, Eq.~(\ref{SL}) becomes
\begin{equation}
\left(\frac{d}{dr_{*}}+W\right)\left(\frac{d}{dr_{*}}-W\right)\tilde{\Phi}_{\rm{II}}=0\label{phireg2}\; .
\end{equation}
Defining $H$ as
\begin{equation}
H=\left(\frac{d}{dr_{*}}-W\right)\tilde{\Phi}_{\rm II}\; ,\label{H}
\end{equation}
Eq.~(\ref{phireg2}) transforms into a first-order differential equation
\begin{equation}
\left(\frac{d}{dr_{*}}+W\right)H=0\label{H1}\; .
\end{equation}
The solution of Eq.~(\ref{H1}) is given
\begin{equation}
H=B_{\rm{II}}\left(\frac{1+\sqrt{f}}{1-\sqrt{f}}\right)^{j+\frac{1}{2}}\left(\frac{j+\frac{1}{2}-\sqrt{f}}{j+\frac{1}{2}+\sqrt{f}}\right)\label{H2}\; .
\end{equation}
Inserting Eq.~(\ref{H2}) into Eq.~(\ref{H}), we will have a first order differential equation of $\tilde\Phi_{\rm{II}}$, and the solution goes like
\begin{equation}
\tilde\Phi_{\rm{II}}=A_{\rm{II}}\left(\frac{1+\sqrt{f}}{1-\sqrt{f}}\right)^{j+\frac{1}{2}}\left(\frac{j+\frac{1}{2}-\sqrt{f}}{j+\frac{1}{2}+\sqrt{f}}\right)
+B_{\rm{II}}\Psi\label{Psi}\; ,
\end{equation}
where $\Psi$ is
\begin{equation}
\Psi=\left(\frac{1+\sqrt{f}}{1-\sqrt{f}}\right)^{j+\frac{1}{2}}\left(\frac{j+\frac{1}{2}-\sqrt{f}}{j+\frac{1}{2}+\sqrt{f}}\right)
\left[\int ^{r}\frac{1}{f}\left(\frac{1-\sqrt{f}}{1+\sqrt{f}}\right)^{2j+1}\left(\frac{j+\frac{1}{2}+\sqrt{f}}{j+\frac{1}{2}-\sqrt{f}}\right)^{2}dr'\right]\; .
\end{equation}

\par In the region where $f(r)\rightarrow1$, Eq.~(\ref{SL}) becomes
\begin{equation}
\frac{d^{2}}{dr^{2}}\tilde{\Phi}_{\rm{ III}}-\left[\frac{\left(j^{2}-\frac{1}{4}\right)}{r^{2}}-\omega^{2}\right]\tilde{\Phi}_{\rm{III}}=0\; .
\end{equation}
Note that in this region $r_{*}\sim r$, the solution will be expressed in terms of Bessel functions as
\begin{equation}
\tilde{\Phi}_{\rm{III}}=A_{\rm{III}}\sqrt{r}J_{j}\left(\omega r\right)
+B_{\rm{III}}\sqrt{r}N_{j}\left(\omega r\right)\; .
\end{equation}
Letting the incoming part of $\tilde{\Phi}_{\rm{ III}}$ to have unit amplitude at $r\rightarrow\infty$, we have the relation between $A_{\rm{III}}$ and $B_{\rm{III}}$ which reads
\begin{equation}
A_{\rm{III}}+i B_{\rm{III}}=\sqrt{2\pi\omega}\; .
\end{equation}

\par Matching the solutions in regions I and II and also the solutions in regions II and III, one can obtain the absorption probability as \cite{unr,junkim}
\begin{equation}
|A_{j}(\omega)|^{2}=4\pi C^{2}\omega^{2j+1}\left(1+\pi C^{2}\omega^{2j+1}\right)^{-2}\approx 4\pi C^{2}\omega^{2j+1}\; ,
\end{equation}
for small $\omega$, where
\begin{eqnarray}
C=\frac{1}{2^{2j+1}\Gamma(j+1)}\left(\frac{j+\frac{3}{2}}{j-\frac{1}{2}}\right)\; ,
\end{eqnarray}
and $\Gamma$ is the gamma function.

\par To evaluate the absorption probabilities for the whole energy range we adopt the WKB approximation. To implement the method it is convenient to change variable to $x=\omega r$ and to take $Q\left(x\right)=\omega^{2}-V$, where Eq.~(\ref{SL}) then becomes
\begin{equation}
\left(\frac{d^{2}}{dx_{*}^{2}}+Q\right)\tilde{\Phi}_{1}=0\; .\label{SL2}
\end{equation}
For energy $\omega^{2}\ll V$ the low energy absorption probabilities are given by the first order WKB approximation, which reads \cite{cholin}
\begin{equation}
|A_{j}\left(\omega\right)|^{2}={\rm exp}\left[-2\int_{x_{1}}^{{x_2}}\frac{dx'}{f(x')}\sqrt{-Q(x')}\right]\label{APLE}\; ,
\end{equation}
where $x_{1}$ and $x_{2}$ are turning points where $Q\left(x_{1},x_{2}\right)=0$ or $V\left(x_{1},x_{2}\right)=\omega^{2}$, for a given energy $\omega$ with potential $V$.

\par When $\omega^{2}\thickapprox V$ the formula in Eq.~(\ref{APLE}) will no longer be appropriate as the exponential goes to infinity.  In this energy regime a suitable method to derive the absorption probabilities is by using the third order WKB approximation of Iyer and Will \cite{iyewil}.  Using the same notation as in Ref. \cite{chocor5}, the absorption probability can be expressed as
\begin{equation}
|A_{j}\left(\omega\right)|^{2}=\frac{1}{1+e^{2S(\omega)}}\; ,\label{3orderWKB}
\end{equation}
where
\begin{eqnarray}
S(\omega)&=&\pi k^{1/2}\left[\frac{1}{2}z_{0}^{2}+\left(\frac{15}{64}b_{3}^{2}-\frac{3}{16}b_{4}\right)z_{0}^{4}\right]\nonumber\\
&&+\pi k^{1/2}\left[+\left(
\frac{1155}{2048}b_{3}^{4}-\frac{315}{256}b_{3}^{2}b_{4}+\frac{35}{128}b_{4}^{2}+\frac{35}{64}b_{3}b_{5}-\frac{5}{32}b_{6}\right)z_{0}^{6}\right]\nonumber\\
&&+\pi k^{-1/2}\left[\left(\frac{3}{16}b_{4}-\frac{7}{64}b_{3}^{2}\right)\right]\nonumber\\
&&-\pi k^{-1/2}\left[\left(\frac{1365}{2048}b_{3}^{4}
-\frac{525}{256}b_{3}^{2}b_{4}+\frac{85}{128}b_{4}^{2}+\frac{85}{128}b_{4}^{2}+\frac{95}{64}b_{3}b_{5}
-\frac{25}{32}b_{6}\right)z_{0}^{2}\right]\; ,
\end{eqnarray}
where $z_{0}^{2}$, $b_n$,  and $k$ are defined by the components of the Taylor series in expanding $Q(x)$ near its peak at $x_{0}$,
\begin{eqnarray}
Q&=&Q_{0}+\frac{1}{2}Q_{0}^{\prime\prime}z^{2}+\sum_{n=3}\frac{1}{n!}\left(\frac{d^{n}Q}{dx^{n}}\right)_{0}z^{n}\nonumber\\
&\equiv &k\left[z^{2}-z_{0}^{2}+\sum_{n=3}b_{n}z^{n}\right],
\end{eqnarray}
and
\begin{eqnarray}
&&z=x-x_{0}\ \ \ ;\ \ \ z_{0}^{2}\equiv -2\frac{Q_{0}}{Q_{0}^{\prime\prime}}\nonumber\\
&&k\equiv\frac{1}{2}Q_{0}^{\prime\prime}\ \ \ ;\ \ \ b_{n}\equiv\left(\frac{2}{n!Q_{0}^{\prime\prime}}\right)\left(\frac{d^{n}Q}{dx^{n}}\right)_{0}.
\end{eqnarray}
The subscript 0 represents the maximum of $Q$ and the primes denote derivatives.

\par
\begin{figure}[!]
\centering
\includegraphics[height=80mm]{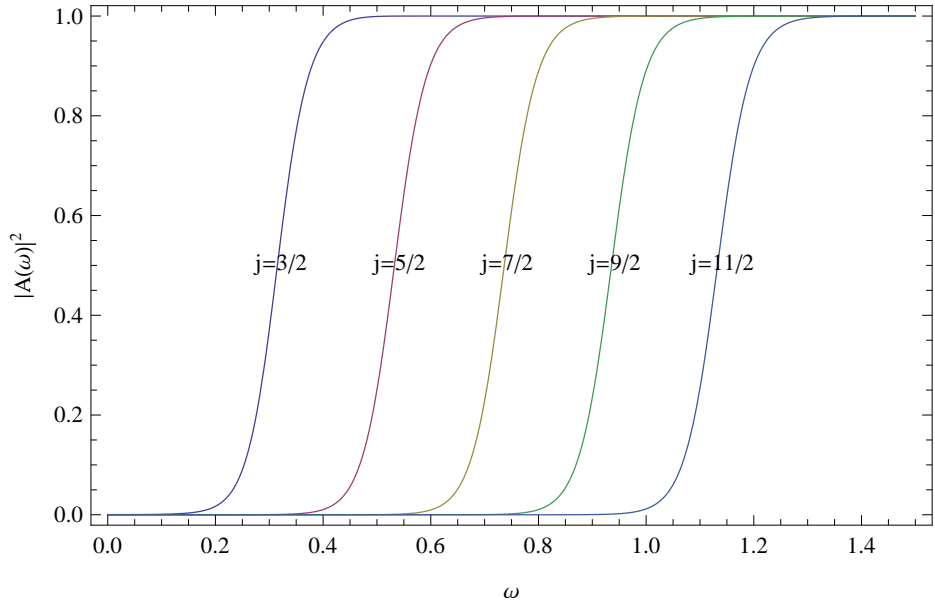}
\caption{\it Absorption probabilities by the WKB approximation for $j=3/2$ to $11/2$.}
\label{fig2}
\end{figure}

 The absorption probabilities given by Eq.~(\ref{3orderWKB}) from $j=3/2$ to $11/2$, and are presented in Fig.~\ref{fig2}. They are plotted as a function of energy $\omega$.  As the parameter $j$ is increased from $3/2$ to $11/2$, the absorption probabilities shift to the right, and all the absorption probabilities approach $1$ as the energy becomes large, that is, in the high energy regime, as expected.

%
%
%

\section{Conclusions and Discussions}\label{sec:conclusion}

\par In this paper we have introduced a novel method to deal with the spin-3/2, or the gravitino, fields in spherically symmetric spacetimes. Our method can be readily extended to higher dimensions, in contrast to the Newman-Penrose formalism which is more specific to four dimensions. In this method we considered the eigenspinor-vectors on higher-dimensional spheres. Actually the eigenspinor-vectors of the Dirac operator in higher-dimensional maximally symmetric spaces is an interesting topic of its own. Along the same lines as the work done by Camporesi and Higuchi \cite{camhig}, for the case of the eigenspinors, one can obtain the spinor-vector eigenfunctions, their eigenvalues and degeneracies on spheres as well as on hyperbolic spaces. Work in this direction is in progress.

\par With the help of these eigenspinor-vectors it is possible to consider spherically symmetric spacetimes in higher dimensions. As we have mentioned earlier, with the addition of the spin-3/2 fields, we can establish the QNM frequencies and the greybody factors in Hawking radiation processes for all spins in higher dimensional spherically symmetric black holes. This will be helpful in understanding the production and the evolution of possible collider produced TeV black holes \cite{kan}.

\par Other than the higher dimensional Schwarzschild black holes one could apply our method to other spherically symmetric spacetimes like the Reissner-Nordstrom as well as the AdS- and dS-black holes. For the charged black hole cases, especially the extremal ones, we could examine the behaviors of supersymmetric black holes. For example, in \cite{cho1} it is found that the asymptotic QNM frequencies are the same for spin-0, 1/2, 1, 3/2 and 2 fields in the extremal case in four dimensions. It is therefore interesting to see if this holds for extremal black holes in higher dimensions.

\par Our method should also be useful in the study of properties of AdS black holes. These will be important in relation to the AdS/CFT correspondence \cite{ahagub} to understand various finite temperature behaviors of the boundary conformal field theory \cite{horhub}. We can examine the gravitino perturbations for the whole space outside the event horizon \cite{liuzay}, not just for the near horizon region. We shall also address this problem in a future work.

%
%

\acknowledgments

CHC and HTC are supported in part by the Ministry of Science and Technology, Taiwan, ROC under the Grants No.~NSC102-2112-M-032-002-MY3, and by the National Center for Theoretical Sciences (NCTS). HTC would like to thank the hospitality of Prof.~Kin-Wang Ng and the Theory Group of the Institute of Physics at the Academia Sinica, Republic of China, where part of this work was done. ASC is grateful to research supported in part by the National Research Foundation of South Africa (Grant No: 91549).

%
%

\appendix

\section{Eigenspinors and eigenspinor-vectors on the 2-sphere}

\par The metric of a 2-sphere can be taken as
\begin{equation}
ds^{2}=\sin^{2}\!\theta_{2}\, d\theta_{1}^{2}+d\theta_{2}^{2}\; ,
\end{equation}
where the zweibein are defined as
\begin{eqnarray}
{e_{\mu}}^{a}={\rm diag}(\sin\theta_{2},1)\ \ \ ;\ \ \ {e_{a}}^{\mu}={\rm
diag}(1/\sin\theta_{2},1)\; .
\end{eqnarray}
Note that we have omitted the overbars for simplicity in this appendix.

\par The Dirac operator is given by
\begin{equation}
\gamma^{a}\nabla_{a}\psi
=\gamma^{a}{e_{a}}^{\mu}\left(\partial_{\mu}+\Gamma_{\mu}\right)\psi\; .
\end{equation}
In this metric the non-vanishing Christoffel symbols $\Gamma^{\alpha}_{\mu\nu}$ and spin connections $\Gamma_{\mu}$ are
\begin{eqnarray}
\Gamma^{\theta_{1}}_{\theta_{1}\theta_{2}}=\cot\theta_{2}\ \ \ ;\ \ \
\Gamma^{\theta_{2}}_{\theta_{1}\theta_{1}}=-\sin\theta_{2}\,\cos\theta_{2}\; , 
\end{eqnarray}
and
\begin{eqnarray}
\Gamma_{\theta_{1}}=\frac{i}{2}\cos\theta_{2}\,\sigma^{3}\ \ \ ;\ \ \
\Gamma_{\theta_{2}}=0 \; .
\end{eqnarray}
The Dirac matrices are chosen to be
\begin{equation}
\gamma^{\theta_{1}}=-\left(\frac{1}{\sin\theta_{2}}\right)\sigma^{2}\ \ \ ;\ \ \ \gamma^{\theta_{2}}=\sigma^{1}\; ,
\end{equation}
where $\sigma^{i}$ are the Pauli matrices.

\par Suppose we write the Dirac operator eigenspinor equation as
\begin{equation}
\gamma^{\mu}\nabla_{\nu}\psi_{\lambda}=i\lambda\psi_{\lambda}
\Rightarrow\left[-\sigma^{2}\frac{1}{\sin\theta_{2}}\partial_{\theta_{1}}+\sigma^{1}\left(\partial_{\theta_{2}}+\frac{1}{2}\cot\theta_{2}\right)
\right]\psi_{\lambda}=
i\lambda\psi_{\lambda}\; ,
\end{equation}
where $i\lambda$ is the eigenvalue. Consider first the eigenvalue equation for $\partial_{\theta_{1}}$,
\begin{equation}
\partial_{\theta_{1}}\chi_{m}=im\chi_{m}\; ,
\end{equation}
that is, $\chi_{m}\sim e^{im\theta_{1}}$. Note that for spinors one should get a sign change for a 2$\pi$ rotation in $\theta_{1}$. Therefore, the eigenvalues of $m$ should be half-integers,
\begin{equation}
m=\pm\frac{1}{2},\pm\frac{3}{2},\pm\frac{5}{2},\cdots.\label{meigenvalue}
\end{equation}

\par Going back to the eigenvalue equation for $\psi_{\lambda}$, we write
\begin{equation}
\psi_{\lambda}=\left(
\begin{array}{c}
A_{\lambda}(\theta_{2})\chi_{m}(\theta_{1}) \\
B_{\lambda}(\theta_{2})\chi_{m}(\theta_{1})
\end{array}
\right).
\end{equation}
Putting this into the eigenvalue equation, we have the following set of equations,
\begin{eqnarray}
&&\left(\partial_{\theta_{2}}+\frac{1}{2}\cot\theta_{2}\right)A_{\lambda}
+\frac{m}{\sin\theta_{2}}A_{\lambda}=i\lambda B_{\lambda}\; ,\nonumber\\
&&\left(\partial_{\theta_{2}}+\frac{1}{2}\cot\theta_{2}\right)B_{\lambda}
-\frac{m}{\sin\theta_{2}}B_{\lambda}=i\lambda A_{\lambda}\; .
\end{eqnarray}
These can be turned into second order equations. For example, for $A_{\lambda}$ we have,
\begin{eqnarray}
\left[\left(\partial_{\theta_{2}}+\frac{1}{2}\cot\theta_{2}\right)
-\frac{m}{\sin\theta_{2}}\right]\left[\left(\partial_{\theta_{2}}+\frac{1}{2}\cot\theta_{2}\right)
+\frac{m}{\sin\theta_{2}}\right]A_{\lambda}=-\lambda^{2}A_{\lambda}\; .\label{Aequation}
\end{eqnarray}
This equation can be further simplified by defining
\begin{equation}
A_{\lambda}=\left(\sin\theta_{2}\right)^{-1/2}u_{\lambda}\; .
\end{equation}
Eq.~(\ref{Aequation}) then becomes
\begin{eqnarray}
&&\frac{d^{2}u_{\lambda}}{d\theta_{2}^{2}}+\left(\lambda^{2}-\frac{m^{2}}{\sin^{2}\theta_{2}}
-\frac{m\cos\theta_{2}}{\sin^{2}\theta_{2}}\right)u_{\lambda}=0\nonumber\\
&\Rightarrow&\frac{d^{2}u_{\lambda}}{d\theta_{2}^{2}}+\left[\frac{1-4\left(m+\frac{1}{2}\right)^{2}}
{16\sin^{2}\frac{\theta_{2}}{2}}+\frac{1-4\left(m-\frac{1}{2}\right)^{2}}
{16\cos^{2}\frac{\theta_{2}}{2}}+\lambda^{2} \right]u_{\lambda}=0\; , 
\end{eqnarray}
with the solution \cite{graryz}
\begin{equation}
u_{\lambda}=\left(\sin\frac{\theta_{2}}{2}\right)^{\alpha+\frac{1}{2}}
\left(\cos\frac{\theta_{2}}{2}\right)^{\beta+\frac{1}{2}}P_{n}^{(\alpha,\beta)}(\cos\theta_{2})\; .
\end{equation}
$P_{n}^{(\alpha,\beta)}(\cos\theta_{2})$ are the Jacobi polynomials with
\begin{equation}
\alpha=\left|m+\frac{1}{2}\right|\ \ \ ;\ \ \
\beta=\left|m-\frac{1}{2}\right|\ \ \ ;\ \ \
n=\pm\lambda-\frac{\alpha+\beta+1}{2}\; .
\end{equation}
$n=0,1,2,\dots$ for $P_{n}^{(\alpha,\beta)}$ to be a polynomial, so that it is convergent in $0\leq\theta_{2}\leq\pi$. Hence, for the values of $m$ given in Eq.~(\ref{meigenvalue}), the eigenvalues of the eigenspinors are
\begin{eqnarray}
\lambda=\pm 1,\pm 2,\pm 3,\dots.\label{eigenspinor}
\end{eqnarray}

\par Next we consider the eigenspinor-vectors of the Dirac operator,
\begin{equation}
\gamma^{\mu}\nabla_{\mu}\psi_{\nu}=i\xi\psi_{\nu}
\end{equation}
on the 2-sphere. From the eigenspinor $\psi_{\lambda}$, one can construct the two independent eigenspinor-vectors on $S^{2}$ using the linear combinations of $\nabla_{\mu}\psi$ and $\gamma_{\mu}\psi$:
\begin{eqnarray}
\psi^{(1)}_{\mu}=\nabla_{\mu}\psi_{\lambda}+a_{1}\gamma_{\mu}\psi_{\lambda}\ \ \ ;\ \ \
\psi^{(2)}_{\mu}=\nabla_{\mu}\psi_{\lambda}+a_{2}\gamma_{\mu}\psi_{\lambda}\; ,
\end{eqnarray}
where
\begin{eqnarray}
&&a_{1}=\frac{i}{2}\left(-\lambda+\sqrt{\lambda^{2}-1}\right)\ \ ;\ \ \xi_{1}=\sqrt{\lambda^{2}-1}\; , \\
&&a_{2}=\frac{i}{2}\left(-\lambda-\sqrt{\lambda^{2}-1}\right)\ \ ;\ \ \xi_{2}=-\sqrt{\lambda^{2}-1}\; .
\end{eqnarray}
They are orthogonal to each other in such a way that
\begin{equation}
\int d\Omega\left(\psi^{(1)}_{\mu}\right)^{\dagger}\psi^{(2)\mu}=0\; .
\end{equation}
Note that they do not obey the condition $\gamma^{\mu}\psi^{(1,2)}_{\mu}=0$. Indeed,
\begin{eqnarray}
\gamma^{\mu}(\nabla_{\mu}\psi_{\lambda}+a_{1,2}\gamma_{\mu}\psi_{\lambda})&=&(i\lambda+2a_{1,2})\psi_{\lambda}\nonumber\\
&=&\pm i\sqrt{\lambda^{2}-1}\,\psi_{\lambda}\; .
\end{eqnarray}
These vanish only for the zero modes, $\xi_{1,2}=0$, which we will exclude. Also, we have
\begin{eqnarray}
\nabla^{\mu}\psi^{(1,2)}_{\mu}&=&\nabla^{\mu}(\nabla_{\mu}\psi_{\lambda}+a_{1,2}\gamma_{\mu}\psi_{\lambda})\nonumber\\
&=&-\frac{1}{2}\sqrt{\lambda^{2}-1}\left(\pm\lambda+\sqrt{\lambda^{2}-1}\right)\psi_{\lambda}\; ,
\end{eqnarray}
which are also non-vanishing.

\par Hence, from the eigenvalues of the eigenspinors in Eq.~(\ref{eigenspinor}), we have the eigenvalues of the eigenspinor-vectors on the 2-sphere,
\begin{eqnarray}
\xi_{1,2}=\pm\sqrt{\lambda^{2}-1}=\pm\sqrt{3},\pm\sqrt{8},\pm\sqrt{15},\dots,
\end{eqnarray}
while the two eigenmodes $\psi_{\mu}^{(1,2)}=\nabla_{\mu}\psi_{\lambda}+a_{1,2}\gamma_{\mu}\psi_{\lambda}$ are linear combinations of $\nabla_{\mu}\psi_{\lambda}$ and $\gamma_{\mu}\psi_{\lambda}$.

%
%

\end{document}